\documentclass[conference]{IEEEtran}
\IEEEoverridecommandlockouts
\usepackage{cite}
\usepackage{amsmath,amssymb,amsfonts}
\usepackage{algorithm}
\usepackage{algpseudocode}
\usepackage{graphicx}
\usepackage{textcomp}
\usepackage{booktabs}
\usepackage{subcaption}
\usepackage{hyperref}\hypersetup{colorlinks=true,linkcolor=blue,filecolor=magenta,urlcolor=cyan,citecolor=blue,}
\usepackage[nolist]{acronym}

\usepackage[table]{xcolor}
\def\BibTeX{{\rm B\kern-.05em{\sc i\kern-.025em b}\kern-.08em
    T\kern-.1667em\lower.7ex\hbox{E}\kern-.125emX}}
\begin{document}

\begin{acronym}[HPC]   
    \acro{GPU}{Graphics Processing Unit}
     \acro{HPC}{High Performance Computing}
    \acro{GEMM}{General Matrix Multiply}
    \acro{LDS}{Local Data Share}
    \acro{BLAS}{Basic Linear Algebra Subprograms}
    \acro{DGEMM}{double-precision GEMM}
    \acro{AI}{Artificial Intelligence}
    \acro{HBM}{High Bandwidth Memory}
    \acro{CRT}{Chinese Remainder Theorem}
    \acro{HPL}{High Performance Linpack}
    \acro{MFMA}{Matrix Fused Multiply-Add}
\end{acronym}

\title{FP64 Performance and the Limits of Ozaki Emulation on HPC Systems\\
}

\author{\IEEEauthorblockN{Katarzyna (Kasia) \'{S}wirydowicz}
\IEEEauthorblockA{\textit{Advanced Micro Devices} \\
Austin, TX \\
kasia.swirydowicz@amd.com}
\and
\IEEEauthorblockN{ Nicholas Malaya}
\IEEEauthorblockA{\textit{Advanced Micro Devices, Inc.} \\
Austin, TX \\
Nicholas.Malaya@amd.com}
\and
\IEEEauthorblockN{ Noah Wolfe}
\IEEEauthorblockA{\textit{Advanced Micro Devices, Inc.} \\
Austin, TX \\
Noah.Wolfe@amd.com}
\and
\IEEEauthorblockN{Paul Bauman}
\IEEEauthorblockA{\textit{Advanced Micro Devices} \\
Austin, TX \\
Paul.Bauman@amd.com}
\and
\IEEEauthorblockN{Pieter Ghysels}
\IEEEauthorblockA{\textit{Advanced Micro Devices, Inc.} \\
Austin, TX \\
Pieter.Ghysels@amd.com}
\and
\IEEEauthorblockN{Jonathan Belof}
\IEEEauthorblockA{\textit{Advanced Micro Devices, Inc.} \\
Austin, TX \\
Jonathan.Belof@amd.com}
}
\maketitle
{\color{black}}
\begin{abstract}
A growing industry narrative holds that native double precision can be deprioritized in \ac{HPC} accelerators, on the hypothesis that low-precision tensor cores can emulate FP64 on demand, and that most \ac{HPC} workloads are bandwidth-bound. This paper evaluates some of those claims and explores the importance of FP64 in scientific computing. We show roofline profiling data of representative \ac{HPC} workloads, across a variety of common computational motifs, to evaluate the arithmetic-intensity and accuracy requirements of these workloads. We also consider FP64 emulation via the Ozaki method to ascertain some of the possibilities and limitations of these approaches. 

These results indicate that Ozaki emulation should not be evaluated as a general replacement for FP64, but as an optimization for the subset of workloads whose dominant computation is sufficiently large, dense, and \ac{GEMM}-shaped. Our analysis indicates this is less than 10\% of the \ac{HPC} ecosystem. Furthermore, even in this favorable regime, standalone \ac{GEMM} speedup is not {\color{black}enough to accelerate (or avoid performance regression) for the application}: numerical behavior, workspace requirements, matrix shape, and integration overhead determine whether the benefit survives at application scale. This paper characterizes those boundaries empirically. Furthermore, we show that dense \ac{GEMM} accounts for only a small fraction of FP64 computation across \ac{HPC} workloads, leaving most application-level FP64 work beyond the reach of \ac{GEMM}-based emulation.
\end{abstract}


\section{Introduction}

Double-precision floating-point arithmetic has been  fundamental to scientific simulation, numerical solvers, and engineering analysis. At the same time, \ac{GPU}s designed primarily for artificial intelligence increasingly devote more resources to low-precision tensor operations and memory capacity while providing comparatively less native FP64 throughput. This trend is often justified by two claims:
that most \ac{HPC} workloads are limited by memory bandwidth, and that low-precision tensor cores can emulate FP64 computation when greater accuracy is required.

These claims conflate several distinct questions. Whether a kernel is compute or bandwidth-bound depends on the balance of the target architecture, not only on the application itself. Moreover, FP64 computation includes vector operations, reductions, stencils, sparse linear algebra, and application-specific kernels, most of which cannot be expressed as dense matrix multiplication. Even when a workload is \ac{GEMM}-dominated, the speed of an emulated \ac{GEMM} does not by itself determine application performance; numerical behavior, workspace requirements, matrix shape, and integration overhead must also be considered.

This paper examines these limitations at both the workload and kernel levels. We first use roofline measurements from 47 configurations spanning eleven \ac{HPC} applications and benchmarks to characterize the importance and computational form of FP64 work (Section~\ref{sec:workloads_results}). Next, we briefly explain the ideas behind two Ozaki schemes (Section~\ref{sec:ozaki}).  We then evaluate Ozaki-style \ac{GEMM} emulation across three current \ac{GPU} architectures, considering accuracy (Section~\ref{sec:accuracy}), performance and memory requirements (Section~\ref{sec:performance}), reproducibility (Section~\ref{sec:robustness}), and sensitivity to matrix scaling and shape. Together, these results identify where low-precision \ac{GEMM} emulation is useful and where substantial native FP64 capability remains necessary.

Real scientific applications present substantially broader requirements than dense matrix multiplication alone, including programmability, performance portability, communication, and application-specific optimization~\cite{bauman2023experiences}. Even \ac{HPL} contains critical components beyond \ac{DGEMM} whose implementation can materially affect performance at scale~\cite{lu2025hpl}.

\section{Prior Work}

The original Ozaki scheme introduced an error-free transformation for expressing high-precision matrix multiplication as a collection of lower-precision matrix products~\cite{ozaki2012error}. Mukunoki et al.~\cite{mukunoki2020dgemm} subsequently adapted the scheme to FP16 Tensor Cores, demonstrating accurate and reproducible \ac{DGEMM} on \ac{GPU}s. Ootomo et al.~\cite{ootomo2024dgemm} extended this approach to INT8 matrix-multiplication units, while Ozaki et al.~\cite{ozaki2025ozaki} introduced the \ac{CRT}-based Ozaki-II scheme. More recent implementations have explored FP8 Tensor Cores and software-emulated FP64 arithmetic~\cite{mukunoki2026fp8}, dynamic selection of the required precision~\cite{schwarz2025guaranteed}, and kernel fusion to reduce intermediate memory traffic~\cite{lu2026emugemm}.

Complementary work has examined the numerical and architectural
limitations of these approaches. Dongarra et al.~\cite{dongarra2024hardware} survey the broader shift toward
reduced- and mixed-precision scientific computing. Abdelfattah et al.~\cite{abdelfattah2025analysis} analyze integer-based matrix multiplication and show that poorly scaled inputs can reduce accuracy or require additional products, while Uchino et
al.~\cite{uchino2026error} provide deterministic error bounds for Ozaki-II. Most recently, a series of papers from Matsuoka~\cite{matsuoka2026fp8} advances the broader claim that FP8-based emulation can replace native FP64 across \ac{HPC} kernels. Our work complements these studies by evaluating both sides of that claim: the accuracy and performance boundaries of standalone Ozaki \ac{GEMM}, and the limited fraction of application-level FP64 computation that is sufficiently large, dense, and \ac{GEMM}-shaped. We additionally quantify the substantial workspace overhead of Ozaki emulation, which can increase the total \ac{HBM} footprint to approximately $3\times$ that of native FP64 \ac{GEMM}, limiting the problem sizes that can reside on the accelerator.

\section{The Prevalence of Matrix Multiplication in the Computational Sciences}
\label{sec:workloads_results}

\begin{figure*}[t]
    \centering
    \includegraphics[width=\textwidth]{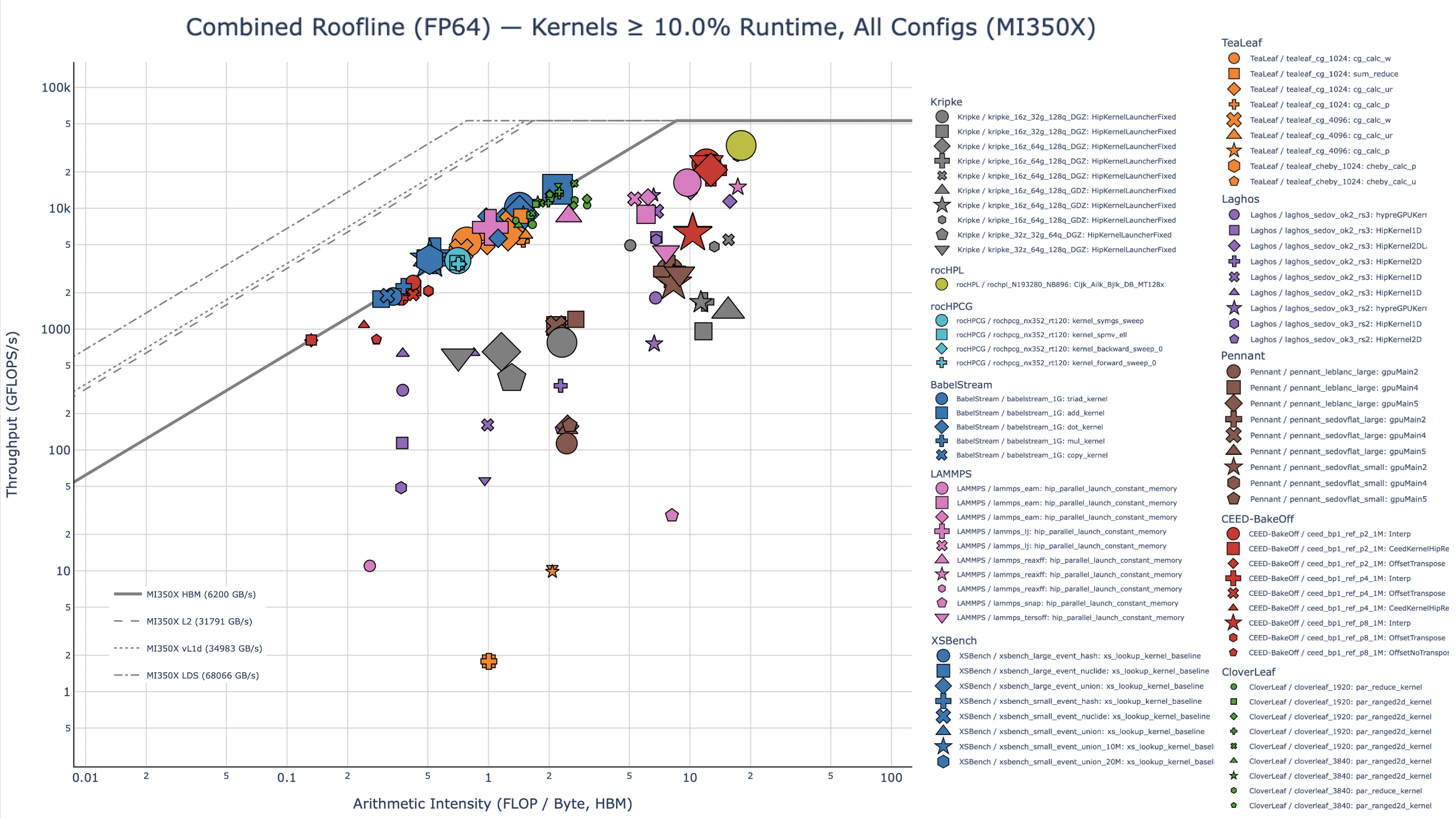}
    \caption{Roofline analysis of 47 workloads spanning eleven \ac{HPC} proxy-application families on AMD Instinct MI350X/MI355X. Each marker represents a measured workload or kernel configuration. The roofline ceilings reflect the attainable throughput imposed by the different levels of the memory hierarchy and by peak FP64 compute throughput. Most workloads are bandwidth-sensitive, while a smaller but important subset approaches the compute ceiling.}
    \label{fig:hpc_roofline}
\end{figure*}

To place \ac{GEMM} emulation in the broader context of \ac{HPC} workloads, Figure~\ref{fig:hpc_roofline} shows the measured FP64 throughput and arithmetic intensity of 47 workloads spanning eleven proxy-application families. 

Our roofline dataset contains 47 workload configurations drawn from 11 applications and benchmarks, comprising 131 plotted kernel instances. Table~\ref{tab:roofline_apps} summarizes the scientific or computational domain represented by each application.
\begin{table}[t]
    \centering
    \caption{Applications and benchmarks represented in the MI350X
    roofline dataset. The dataset comprises 47 application
    configurations and 131 plotted kernel instances.}
    \label{tab:roofline_apps}
    \footnotesize
    \begin{tabular}{@{}p{0.25\linewidth}p{0.65\linewidth}@{}}
        \toprule
        Application & Scientific or computational domain \\
        \midrule
        XSBench &
        Monte Carlo neutron transport and reactor physics \\
        
        TeaLeaf &
        Heat conduction using implicit stencil-based solvers \\
        
        CloverLeaf &
        Compressible Eulerian hydrodynamics \\
        
        CEED Bake-Off &
        High-order finite-element operator evaluation and PDE
        discretization \\
        
        Laghos &
        High-order finite-element Lagrangian hydrodynamics \\
        
        PENNANT &
        Lagrangian hydrodynamics on unstructured meshes \\
        
        LAMMPS &
        Classical molecular dynamics and materials modeling \\
        
        Kripke &
        Deterministic discrete-ordinates particle transport \\
        
        rocHPL &
        Dense linear algebra and system-level FP64 benchmarking \\
        
        roc\ac{HPC}G &
        Sparse iterative linear algebra and system benchmarking \\
        
        BabelStream &
        Device-memory bandwidth benchmarking \\
        \bottomrule
    \end{tabular}
\end{table}

Most workloads lie below the \ac{HBM} ridge point and are therefore primarily bandwidth-sensitive. However, a distinct compute-intensive subset remains sensitive to native FP64 throughput. Moreover, roofline position alone does not establish applicability to Ozaki-style emulation: only workloads whose FP64 computation is dominated by sufficiently large dense GEMMs can directly benefit. Thus, memory bandwidth, native FP64 throughput, and
\ac{GEMM} emulation address complementary rather than interchangeable parts of the \ac{HPC} workload portfolio.


This imbalance may become substantially more pronounced in future accelerators designed primarily for \ac{AI}. NVIDIA's Vera Rubin \ac{GPU} provides 33~TFLOP/s of native FP64 throughput\footnote{NVIDIA separately reports 200~TFLOP/s of FP64 \ac{DGEMM} throughput for Rubin using Tensor Core-based emulation. This rate is specific to \ac{GEMM} and does not represent the native FP64 ceiling for general vector, reduction, sparse, or application-specific kernels.} and 22~TB/s of \ac{HBM}4 bandwidth, while AMD's Instinct MI455X provides only 5~TFLOP/s of FP64 throughput with 23.3~TB/s of \ac{HBM}4 bandwidth~\cite{nvidia2026rubin,amd2026mi455x}. These devices therefore offer considerably less native FP64 compute throughput per unit of memory bandwidth than the MI355X, despite their large increases in memory and low-precision tensor performance.

The corresponding \ac{HBM} ridge points, computed as the ratio of peak FP64 throughput to peak memory bandwidth, are approximately 1.50~FLOP/byte for Rubin and 0.21~FLOP/byte for MI455X. These values are substantially below the MI355X ridge point shown in Figure~\ref{fig:hpc_roofline}. Under a first-order roofline projection, many kernels that are bandwidth-bound on MI355X would consequently become compute-bound on Rubin and MI455X. {\color{black} Approximately 50 of the 131 plotted kernel instances would cross the HBM ridge on Rubin and roughly 100 on MI455X.} Thus, the observation that current \ac{HPC}
workloads are frequently bandwidth-bound does not imply that native FP64 throughput can be reduced without consequence: changing the machine balance changes the bottleneck and would render many existing HPC codes compute-bound. 

For FP64 kernels that cannot be expressed as dense \ac{GEMM}, low-precision tensor-core emulation does not remove this limitation.

\section{Matrix Multiplication and the Ozaki Scheme}

\label{sec:ozaki}
\subsection{Ozaki-I and Ozaki-II emulation}
\ac{GEMM} is a fundamental \ac{BLAS} operation: given matrices $A$ and $B$ with matching dimensions (and an optional $D$), compute $C = AB + D$. {\it Ozaki scheme I}~\cite{ozaki2012error}  and {\it Ozaki Scheme II} \cite{ozaki2025ozaki} are used for emulating double precision \ac{GEMM} using lower precision computing (FP8 or INT8). 

In Ozaki-I scheme, the matrices are iteratively scaled and split into {\it slices} to form $A = \sum_{i=1}^{s_A} \tilde{A}_i$ and $B = \sum_{i=1}^{s_B} \tilde{B}_i$, where $\tilde{A}_i$ and $\tilde{B}_i$ are scaled slices. Slices  without their scaling factors fit into prescribed precision (usually INT8 or FP8, but FP16 can be used). They are multiplied through, scaled back and added together to form $C$. This algorithm requires a maximum of $s_A\cdot s_B$ low-precision GEMMs (some \ac{GEMM}s, which would not contribute to the result can be dropped). One needs 7-8 such slices to preserve double precision accuracy if using INT8 slices {\color{black}  for matrices with large exponent range, see~\cite{ootomo2024dgemm}}.

As mentioned before, Ozaki-II scheme is based on an entirely different principle: it uses \ac{CRT}. Rows of $A$ and columns of $B$ are scaled by a diagonal matrix with powers of $2$ on the diagonal so that the elements are integers; several methods exist for choosing the scaling factors~\cite{ozaki2025ozaki, kawakami2026improved}. After the scaling step, one uses a set of $s$ coprime numbers called {\it moduli} (typically close to the upper limit of INT8, i.e., 255, 253, etc) and computes $s$ slices based on the modulo arithmetic. The "matching" slices (originating from the division by the same number) are multiplied in the modulo arithmetic. The result is reconstructed using Garner's algorithms ({\it \ac{CRT} reconstruction}) and scaled back to FP64. Ozaki-II typically requires less \ac{GEMM}s than Ozaki-I scheme ($s$ for $s$ moduli). It can also be implemented using FP8 for the \ac{GEMM} step~\cite{uchino2026doubleprecisionmatrixmultiplicationemulation}. {\color{black}To preserve FP64 accuracy, one needs 15-16 moduli (depending on the exponent range).}

{\color{black} As can be inferred from the above, Ozaki-II scheme requires less \ac{GEMM}s and is expected to be faster (especially if high number of slices is needed), and thus it immediately gained popularity in \ac{HPC} community, which lead to highly optimized open source implementation. Thus we limit our studies on Ozaki-I to its accuracy.}

\subsection{Terminology}

By {\it FP64 \ac{GEMM}} or {\it native \ac{GEMM}} we understand \ac{GEMM} computed using \ac{BLAS} \texttt{dgemm} function on CPU or \ac{GPU} with input and output matrices stored in double precision accuracy.

We use {\it [format] Ozaki-N k}, where [format] is either FP8 or INT8, $N$ is either I or II and $k=2, \ldots 16$ is the number of slices or moduli, to denote specific Ozaki scheme variants.

\section{Accuracy of standalone emulated \ac{GEMM}}
\label{sec:accuracy}
\subsection{Test setup}
In the set of tests in the current and the next sections we use matrices with tightly controlled {\it condition number}. Let $A$ be an $n\times m$ real-valued matrix. We define the condition number as 
\begin{equation}
\kappa\left(A\right)  = \frac{\sigma_{\text{max}}\left(A\right)}{\sigma_{\text{min}}\left(A\right)},
\end{equation}
where $\sigma_{\text{min}}$ and $\sigma_{\text{max}}$ are the smallest and largest singular values of $A$, respectively. The condition number is an internal property of the matrix; it does not change with the algorithm or floating point precision in which the operations on the matrix are carried out. Condition number can be considered a {\it sensitivity measure}: the higher it is, the more the matrices are sensitive to small perturbation(s) in the computations. 

{\color{black} In this work, we only present results for square matrices, in both accuracy and performance experiments. We studied non-square matrices, however, the performance results depended only on the internal dimension (the dimension along which inner product is performed): this dimension needed to be sufficiently large to see speedups. This aligns with the square matrix results. }

In the first set of tests we generate (pseudo-random) matrices and enforce the condition number, see Algorithm~\ref{alg:cond}. In Algorithm~\ref{alg:cond}, qr command performs QR decomposition using \texttt{dgeqrf} followed by \texttt{dorgqr} to ensure orthonormal columns (which guarantees that both $Q_U$ and $Q_V$ have condition numbers equal $1$). The linspace command in Line 3 generates an array of $n$ evenly spaced real numbers starting at $10^0$ and ending at $10^k$. Random matrices are generated and stored in FP64; the method given in Algorithm~\ref{alg:cond} is executed entirely on the \ac{GPU} by using appropriate library (ROCm or CUDA) functions for QR decomposition and a custom kernel to generate $S$. $A$ in Step 6 is computed by two standard FP64 \texttt{dgemm}s.

\begin{algorithm}
\caption{Algorithm used to generate test matrices}
\label{alg:mm}
\begin{algorithmic}[1]    

\Require Positive integers $m$  and $n$; condition number $10^k$

\State $U = rand(m, n)$ \Comment{mean 0.0, std 1.0}
\State $V = rand(n, n)$ \Comment{mean 0.0, std 1.0}
\State $S = \text{diag}\left(\text{linspace}\left(10^0, 10^k, n\right)\right)$ \Comment{dim $n\times n$}
\State $\left[ Q_U, R_U\right] = \text{qr}(U)$
\State $\left[ Q_V, R_V\right] = \text{qr}(V)$
\State $A = Q_U S Q_V^T $

\State \Return $A$

\end{algorithmic}
\label{alg:cond}
\end{algorithm}

\subsection{Basic tests}

To measure accuracy, we generate a series of matrices with condition numbers ranging from $1e0$ to $1e17$ and multiply them with native FP64 \ac{GEMM}, INT8 Ozaki-II 12 \ac{GEMM}, INT8 Ozaki-II 16 \ac{GEMM}, FP8 Ozaki-II 10 \ac{GEMM}, and FP8 Ozaki-II 12 \ac{GEMM}. We use RIKEN's GEMMul8 code~\cite{gemmul8}, commit \texttt{9c1dba457fd5}. We measure the accuracy with the formula
\begin{equation}
      \varepsilon_F \;=\;
    \frac{\|C_{\mathrm{ozaki}} - C_{\mathrm{FP64}}\|_F}{\|C_{\mathrm{FP64}}\|_F},
\end{equation}
where $\|\cdot\|_F$ denotes Frobenius norm.

\begin{figure}
    \centering
    \includegraphics[width=\linewidth]{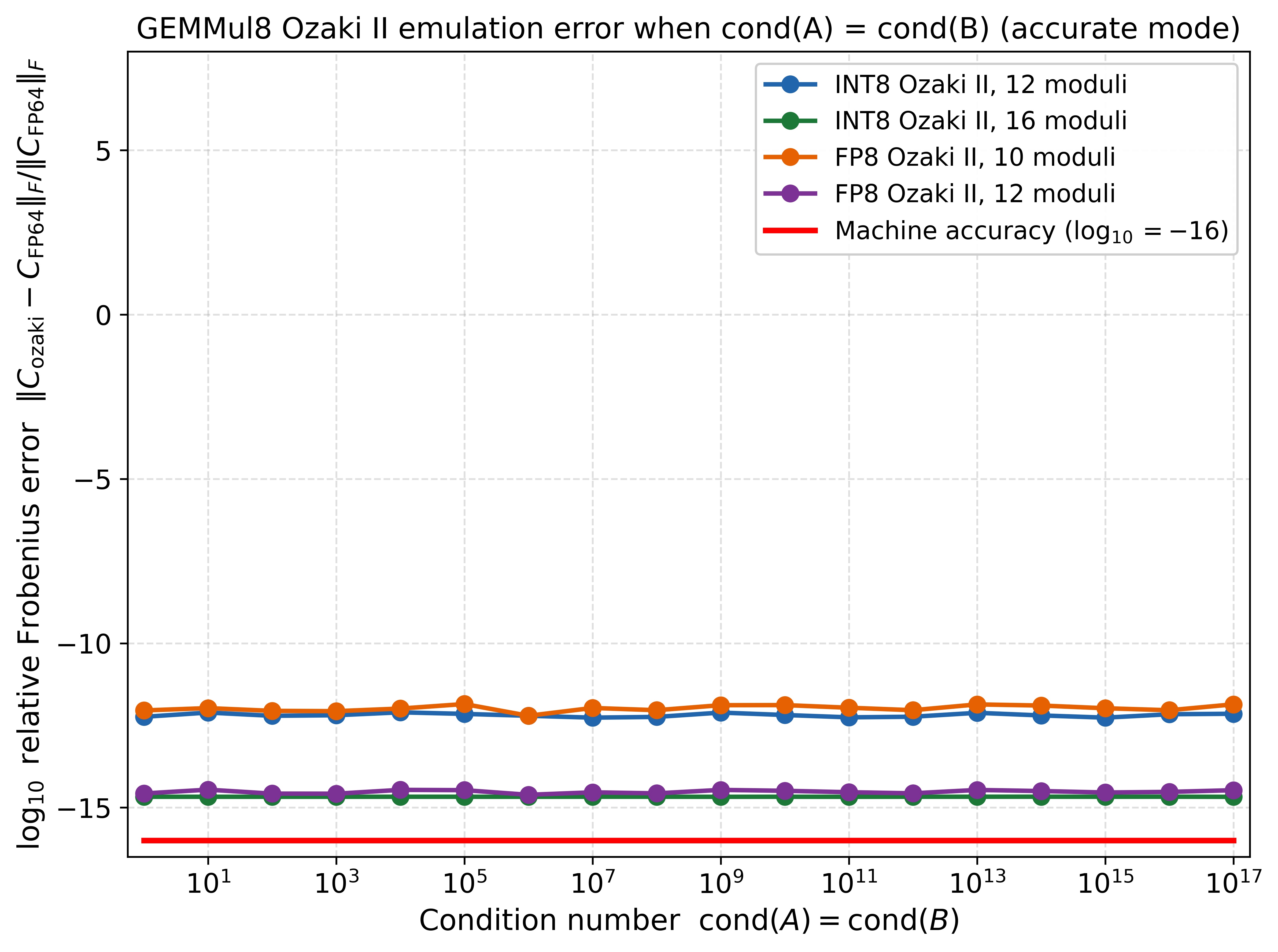}
    \caption{Accuracy of various Ozaki scheme II computations using two (distinct) matrices $A$ and $B$ with prescribed condition number ($D=0$). Matrix size $4096 \times 4096$.}
    \label{fig:ozaki_accuracy}
\end{figure}

The results are shown in Figure~\ref{fig:ozaki_accuracy}. The GEMMul8 code allows for two pathways: accurate (shown in figure) and fast (not shown). The difference lies in how the scaling factors in Ozaki-II are computed; the performance difference is generally within 5\% especially in high moduli count. One can observe that as long as enough moduli is used, \ac{GEMM} accuracy stays consistent and in an acceptable range, even if multiplying two extremely ill-conditioned matrices. Using more moduli does not improve the accuracy any further. In further experiments, we used various matrix shapes and sizes, but we omit the plots, as they look analogous. In this simple test, ill-conditioning does not {\it break} Ozaki scheme \ac{GEMM} - the results are independent on the condition number and only depend on the number of moduli.   

\begin{figure}
    \centering
    \includegraphics[width=\linewidth]{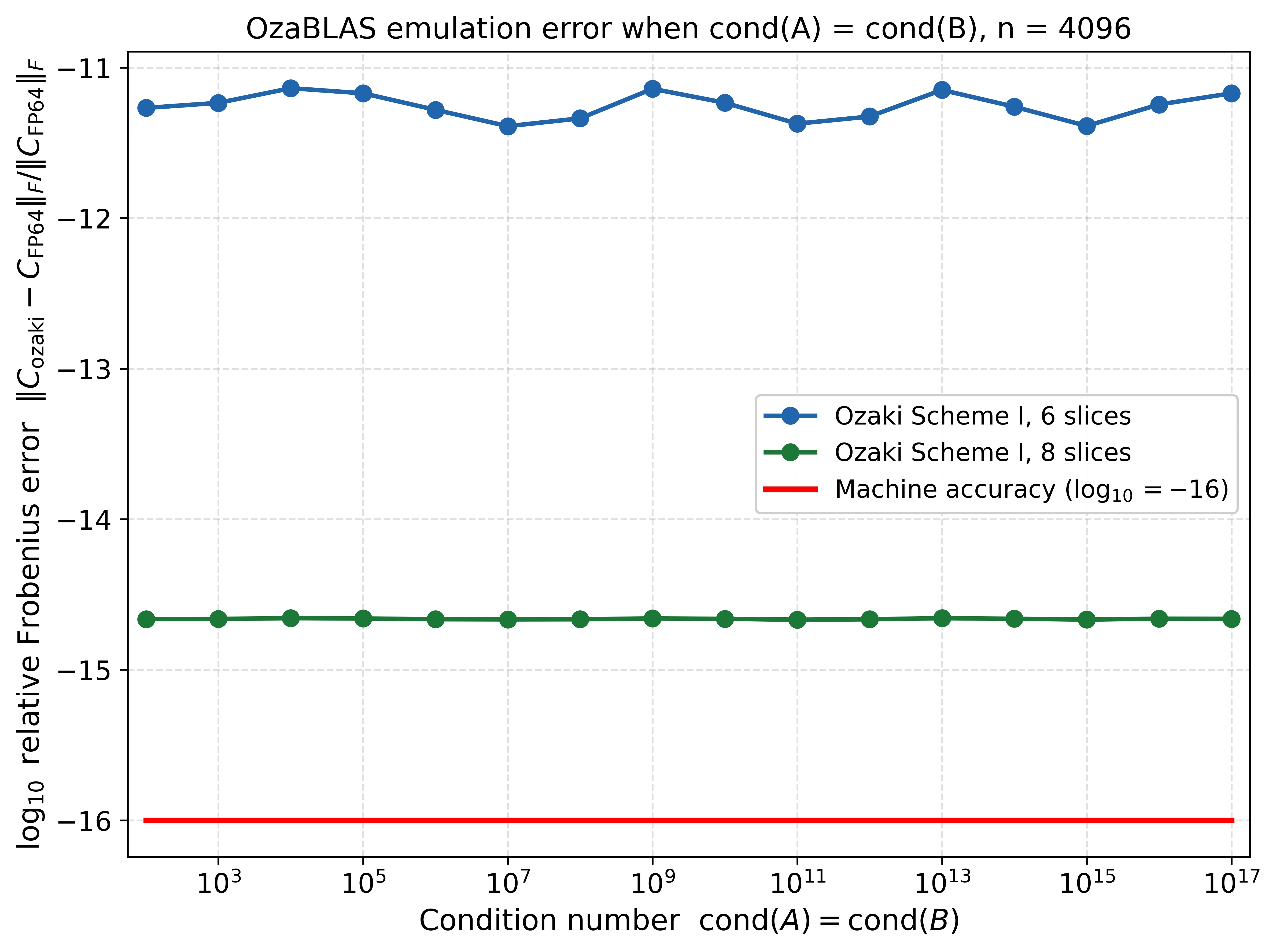}
    \caption{Accuracy of Ozaki I scheme computations using two (distinct) matrices $A$ and $B$ with prescribed condition number ($D=0$). Matrix size $4096 \times 4096$. The 6 and 8 slice results differ significantly from the IEEE machine accuracy.}
    \label{fig:ozaki1_accuracy}
\end{figure}

We do an analogous test using Ozaki I scheme. In this case we use ozablas code~\cite{ozablas}, commit \texttt{043719e}. Note that ozablas implements Ozaki I with INT8 backend and no FP8 backend is available. In case of Ozaki I, less slices are needed; using more slices than shown in Figure~\ref{fig:ozaki1_accuracy} does not improve the accuracy any further. The conclusions are identical: the accuracy depends on the number of slices and not on the condition number, 

These tests do not prove that Ozaki scheme I and Ozaki scheme II themselves are accurate -- if the matrix condition number is as large as 1e17, the FP64 results cannot be trusted -- they might be very different from the exact ("pen and paper") analytic result. The tests we performed simply show that native and emulated \ac{GEMM} produce very similar outputs for similar inputs.
\subsection{Test that breaks the accuracy}
To further stress-test emulation we perform another experiment, based on~\cite{demmel2024presentation}. We generate random matrices $A$ and $B$ with mean value of $0$ and standard deviation of $1$, each matrix of size $1024 \times 1024$. Next, we scale the matrices: $A_s = AD$ and $B_s = D^{-1}B$, where $D$ is a diagonal matrix with entries $2^t$ with $t\in [-m, m]$ is an integer chosen randomly and $m = 0, 2, \ldots, 54$. In exact arithmetic, $AB = A_sB_s$. We compute six outputs, as follows:
\begin{align}
C_1 &= AB \quad \text{in FP64}, \\
C_2 &= A_sB_s \quad \text{in FP64}, \\
C_3 &= A_sB_s \quad \text{using INT8 Ozaki II-16},\\
C_4 &= AB \quad \text{using INT8 Ozaki II-16}, \\
C_5 &= A_sB_s \quad \text{using FP8 Ozaki II-12}, \\
C_6 &= AB \quad \text{using FP8 Ozaki II-12}.
\end{align}

\begin{figure}
    \centering
    \includegraphics[width=\linewidth]{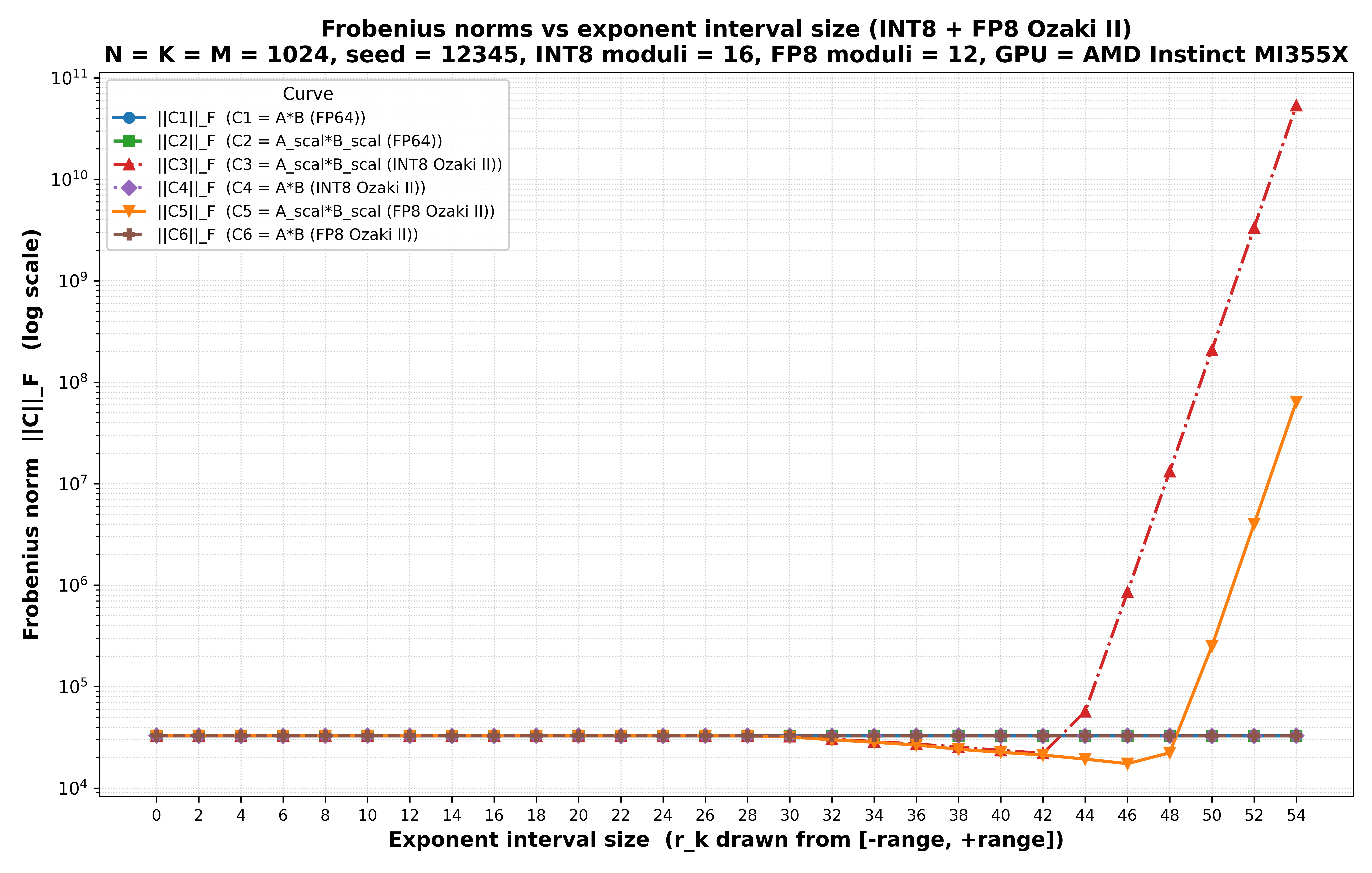}
    \caption{Results of the scaling test using GEMMul8 code.}
    \label{fig:ozaki_fail_norms}
\end{figure}

\begin{figure}
    \centering
    \includegraphics[width=\linewidth]{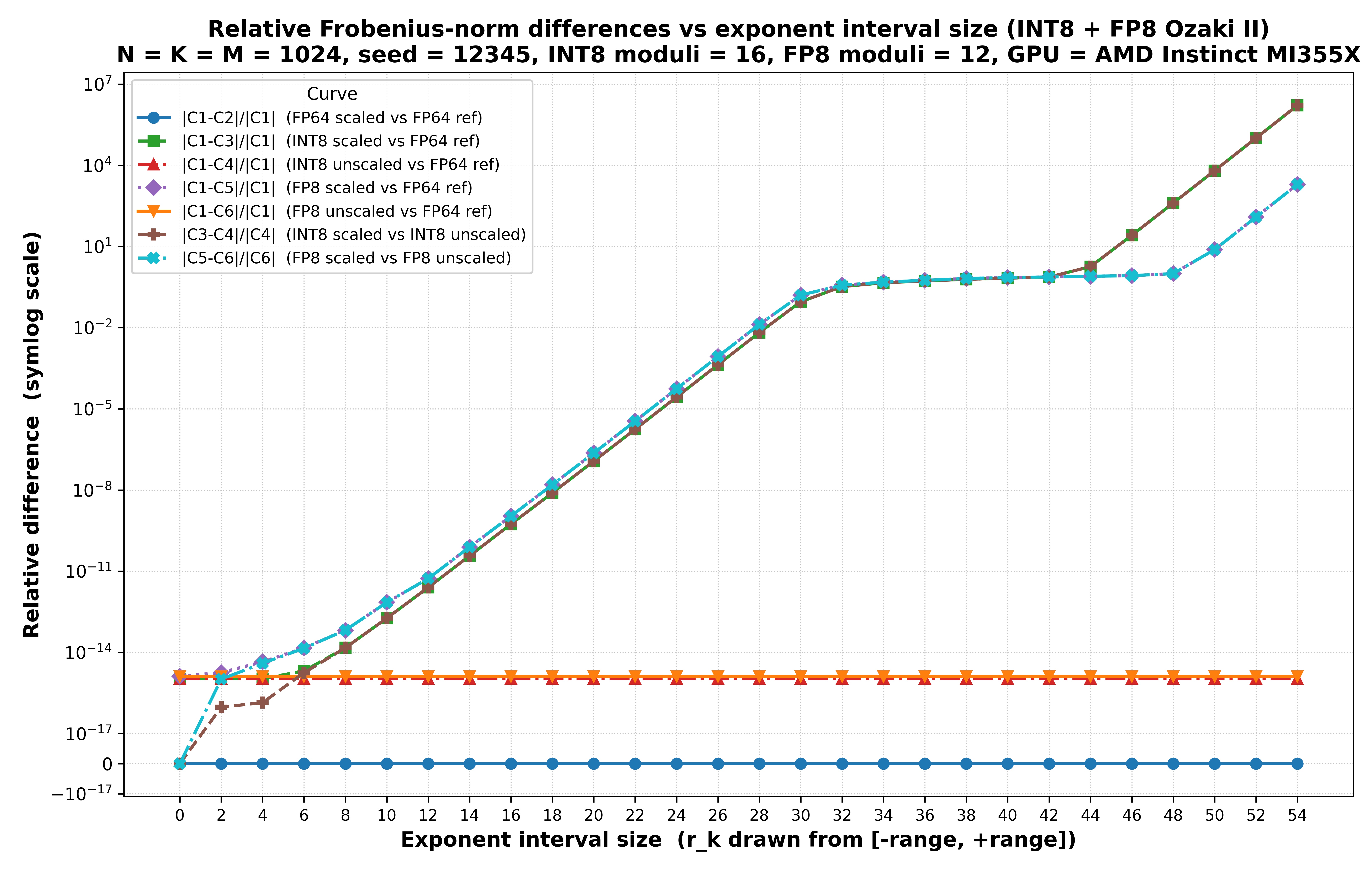}
    \caption{Pairwise Frobenius-norm differences under the product-preserving scaling $A_s=AD$ and $B_s=D^{-1}B$. Native FP64 remains invariant, whereas Ozaki-II becomes increasingly sensitive as the operand exponent range $m$ grows. Thus, algebraically equivalent matrix products can produce materially different Ozaki II results solely because of operand scaling.}
    \label{fig:ozaki_fail_comp}
\end{figure}

We compare the norms of $C_i$s on Figure~\ref{fig:ozaki_fail_norms}. It is clear from the graph that both $C_3$ and $C_5$ start differing from the correct result ($C_1$) at $m \approx 34$. The norm deviates more severely if INT8 arithmetic is used (line with red triangles). In 
Figure~\ref{fig:ozaki_fail_comp} we compare the resulting matrices pair-wise. If $C_3$ or $C_5$ is used, the difference eventually becomes very large; even modest $m$ decreases accuracy. At the same time, there is no difference between $C_1$ and $C_2$ (flat line at $y=0$) and $C_4$ and $C_6$ are good approximations of $C_1$. This data shows that Ozaki scheme II is sensitive to scaling even if we use a large number of slices.

\section{Performance of standalone emulated \ac{GEMM}}
\label{sec:performance}

In our next set of tests, we evaluate the performance. We use three computational systems, see Table~\ref{tab:systems} for details. 

\begin{table}[t]
\caption{System configurations.}
\label{tab:systems}
\centering
\scriptsize
\rowcolors{2}{gray!10}{white}
\begin{tabular}{llll}
\toprule
\rowcolor{gray!25}
\textbf{Name} & \textbf{\ac{GPU}} & \textbf{CPU} & \textbf{Software Stack} \\
\midrule
MI355X & AMD MI355X & AMD EPYC 9575F & ROCm 7.2.1, GCC 13.3.0 \\
B200   & NVIDIA B200 & Intel Xeon 6960P & CUDA 13.3.73, GCC 13.3.0 \\
GB300  & NVIDIA GB300 & Arm Neoverse-V2 & CUDA 13.1, GCC 13.3.0 \\
\bottomrule
\end{tabular}
\end{table}

According to the theoretical specifications, on Nvidia B200 FP8 and INT8 \ac{GEMM} achieves up to 4.5 POPS (Peta OPS) and FP64 \ac{GEMM} achieves 40 TFLOPs, which gives a speedup of $112.5 \times$ per single \ac{GEMM}. Hence, we can expect speedup up to $7.04\times$ for INT8 Ozaki-II 16 and up to $9.38\times$ for FP8 Ozaki-II 12. On B300, we get {\color{black}only} 1.3 TFLOPS for FP64 \ac{GEMM} {\color{black} according to the technical specs~\cite{nvidia2026GB300}}, 10 PFLOPS per FP8 and 330 TOPs for INT8. The ratio per single \ac{GEMM} is $254\times$ for INT8 and $7692 \times$ for FP8, which gives up to $16\times$ speedup for INT8 Ozaki-II 16 and up to $641\times$ for FP8 Ozaki-II 12.
On MI355X, the theoretical peak FP64 matrix performance is 78.6 TFLOPS, the peak INT8 is 5 POPs. and peak FP8 is 5 PFlops. This leads to $63.6$ ratio, which caps the speedup at $3.98$ for INT8 Ozaki-II 16. 

These numbers are not achievable in practice (Ozaki II does more than just GEMMs: the matrices need to be preprocessed and the results need to be assembled, and there is memory traffic and cost of kernel launches) but they constitute the theoretical upper limit.

Results are shown in Tables~\ref{tab:ab_size_ozaki_MI355x}, ~\ref{tab:ab_size_ozaki_B200} and~\ref{tab:ab_size_ozaki_GB300}. The workspace size was measured with GEMMul8's own function \texttt{gemmul8::workSize}. Note that for the $40960\times 40960$ matrix we got an out-of-memory error on B200.

Here is what we can infer from the data:

\begin{itemize}
\item[1.] INT8 Ozaki II scheme is faster than native \ac{GEMM} only for large enough matrices ($4096 \times 4096$ and larger on AMD MI355X and Nvidia B200). At this size, one needs $\sim 850$ MB (for Ozaki II-16) of workspace storage. 
\item[2.] The workspace size grows with the matrix size; for instance, to store $40960 \times 40960$ $A$ and $B$ (and resulting $C$) we need about 37.5 GB. However, the workspace size (for INT8 Ozaki-II 16) increases memory requirement three-fold (to about 120 GB). The increase is only larger for FP8 Ozaki II.
\item[3.] The speedup achieved by Ozaki is larger with larger matrix size. \ac{GEMM} performs $\text{O}\left(n^3\right)$ operations (where $n$ is the matrix size); all the supplementary operations scale with $n^2$, hence at large values of $n$ GEMMs dominate over everything else, including a constant cost of kernel launches that limits the performance at small matrix sizes.
\item[4.] While FP8 Ozaki-II 12 leads to about $5\times$ higher speedup (compared to INT8 Ozaki II-16) on GB300, almost $2\times$ as much additional memory is required.
\item[5.] On GB300, INT8 Ozaki II never reaches performance of native FP64 delivered by B200. FP8 Ozaki needs very large ($8192 \times 8192$ or larger) matrix to reach it.
\item[6.] The speedups (emulated \ac{GEMM} vs native \ac{GEMM}) are very high on GB300 mostly because the native FP64 performance is greatly reduced. 
\end{itemize}

\begin{table*}[htbp]
  \centering
  \caption{AMD MI355X: INT8 GEMMul8 Ozaki~II performance and workspace for $C = AB$ with $A \neq B$ (both square, same size) with condition number $\kappa = 10^{3}$. Performance is the mean of 50 timed runs after 10 warmup iterations.}
  \label{tab:ab_size_ozaki_MI355x}
  \begin{tabular}{rrrrrrrr}
    \toprule
    & Native & \multicolumn{3}{c}{Ozaki~II (12 moduli)} & \multicolumn{3}{c}{Ozaki~II (16 moduli)} \\
    \cmidrule(lr){3-5} \cmidrule(lr){6-8}
    Size & TFLOP/s & TFLOP/s & WS (MB) & Speedup & TFLOP/s & WS (MB) & Speedup \\
    \midrule
    384 & 3.94 & 0.44 & 40.1 & 0.11$\times$ & 0.39 & 42.6 & 0.10$\times$ \\
    512 & 9.64 & 1.09 & 41.8 & 0.11$\times$ & 0.96 & 44.8 & 0.10$\times$ \\
    1024 & 38.33 & 7.79 & 71.0 & 0.20$\times$ & 6.66 & 83.0 & 0.17$\times$ \\
    2048 & 58.53 & 38.48 & 188.0 & 0.66$\times$ & 30.76 & 236.0 & 0.53$\times$ \\
    4096 & 65.54 & 86.06 & 656.0 & 1.31$\times$ & 66.21 & 848.0 & 1.01$\times$ \\
    8192 & 68.61 & 131.85 & 2560.0 & 1.92$\times$ & 102.55 & 3328.0 & 1.49$\times$ \\
    16384 & 69.13 & 161.49 & 10240.1 & 2.34$\times$ & 123.88 & 13312.1 & 1.79$\times$ \\
    24576 & 65.76 & 173.41 & 23040.1 & 2.64$\times$ & 129.93 & 29952.1 & 1.98$\times$ \\
    32768 & 66.26 & 176.10 & 40960.1 & 2.66$\times$ & 134.36 & 53248.1 & 2.03$\times$ \\
    40960 & 64.36 & 183.51 & 64000.2 & 2.85$\times$ & 136.31 & 83200.2 & 2.12$\times$ \\
    \bottomrule
  \end{tabular}
\end{table*}

The workspace reported in Table~\ref{tab:ab_size_ozaki_MI355x} is not a minor scratch allocation: it consumes a substantial fraction of GPU HBM. For the $40960\times40960$ case, storing $A$, $B$, and $C$ requires approximately 37.5~GB, while Ozaki-II requires an additional 64.0~GB with 12 moduli or 83.2~GB with 16 moduli. The resulting total footprint is therefore approximately 100--120~GB, or $2.6$--$3.1\times$ that of
native FP64 GEMM.

\begin{table*}[htbp]
  \centering
  \scriptsize
  \caption{Nvidia B200: INT8 and FP8 GEMMul8 Ozaki~II performance and workspace for $C = AB$ with $A \neq B$ (both square, same size) with condition number $\kappa = 10^{3}$. Performance is the mean of 50 timed runs after 10 warmup iterations.}
  \label{tab:ab_size_ozaki_B200}
  \begin{tabular}{rrrrrrrrrrrrrr}
    \toprule
    & Native & \multicolumn{3}{c}{INT8 Ozaki~II (12 moduli)} & \multicolumn{3}{c}{INT8 Ozaki~II (16 moduli)} & \multicolumn{3}{c}{FP8 Ozaki~II (10 moduli)} & \multicolumn{3}{c}{FP8 Ozaki~II (12 moduli)} \\
    \cmidrule(lr){3-5} \cmidrule(lr){6-8} \cmidrule(lr){9-11} \cmidrule(lr){12-14}
    Size & TFLOP/s & TFLOP/s & WS (MB) & Speedup & TFLOP/s & WS (MB) & Speedup & TFLOP/s & WS (MB) & Speedup & TFLOP/s & WS (MB) & Speedup \\
    \midrule
    384 & 5.65 & 0.61 & 40.1 & 0.11$\times$ & 0.15 & 42.6 & 0.03$\times$ & 0.47 & 48.1 & 0.08$\times$ & 0.42 & 51.5 & 0.07$\times$ \\
    512 & 11.77 & 1.40 & 41.8 & 0.12$\times$ & 1.21 & 44.8 & 0.10$\times$ & 1.11 & 51.5 & 0.09$\times$ & 1.01 & 55.5 & 0.09$\times$ \\
    1024 & 27.10 & 9.35 & 71.0 & 0.34$\times$ & 7.74 & 83.0 & 0.29$\times$ & 6.79 & 110.0 & 0.25$\times$ & 6.07 & 126.0 & 0.22$\times$ \\
    2048 & 34.20 & 39.57 & 188.0 & 1.16$\times$ & 31.30 & 236.0 & 0.92$\times$ & 25.88 & 344.0 & 0.76$\times$ & 22.40 & 408.0 & 0.66$\times$ \\
    4096 & 35.18 & 96.22 & 656.0 & 2.74$\times$ & 73.79 & 848.0 & 2.10$\times$ & 52.03 & 1280.0 & 1.48$\times$ & 44.05 & 1536.0 & 1.25$\times$ \\
    8192 & 35.64 & 147.96 & 2560.0 & 4.15$\times$ & 109.42 & 3328.0 & 3.07$\times$ & 67.64 & 5120.0 & 1.90$\times$ & 57.05 & 6144.0 & 1.60$\times$ \\
        16384 & 35.94 & 174.70 & 10240.1 & 4.86$\times$ & 131.10 & 13312.1 & 3.65$\times$ & 75.82 & 20480.1 & 2.11$\times$ & 63.72 & 24576.1 & 1.77$\times$ \\
    24576 & 36.41 & 152.20 & 23040.1 & 4.18$\times$ & 119.49 & 29952.1 & 3.28$\times$ & 78.62 & 46080.1 & 2.16$\times$ & 65.95 & 55296.1 & 1.81$\times$ \\
    32768 & 36.11 & 161.46 & 40960.1 & 4.47$\times$ & 126.60 & 53248.1 & 3.51$\times$ & 79.66 & 81920.1 & 2.21$\times$ & 66.77 & 98304.1 & 1.85$\times$ \\
    \bottomrule
  \end{tabular}
\end{table*}

\begin{table*}[htbp]
  \centering
  \scriptsize
  \caption{Nvidia GB300: INT8 and FP8 GEMMul8 Ozaki~II performance and workspace for $C = AB$ with $A \neq B$ (both square, same size) with condition number $\kappa = 10^{3}$. Performance is the mean of 50 timed runs after 10 warmup iterations.}
  \label{tab:ab_size_ozaki_GB300}
  \begin{tabular}{rrrrrrrrrrrrrr}
    \toprule
    & Native & \multicolumn{3}{c}{INT8 Ozaki~II (12 moduli)} & \multicolumn{3}{c}{INT8 Ozaki~II (16 moduli)} & \multicolumn{3}{c}{FP8 Ozaki~II (10 moduli)} & \multicolumn{3}{c}{FP8 Ozaki~II (12 moduli)} \\
    \cmidrule(lr){3-5} \cmidrule(lr){6-8} \cmidrule(lr){9-11} \cmidrule(lr){12-14}
    Size & TFLOP/s & TFLOP/s & WS (MB) & Speedup & TFLOP/s & WS (MB) & Speedup & TFLOP/s & WS (MB) & Speedup & TFLOP/s & WS (MB) & Speedup \\
    \midrule
    384 & 0.49 & 0.41 & 40.1 & 0.85$\times$ & 0.35 & 42.6 & 0.73$\times$ & 0.44 & 48.1 & 0.90$\times$ & 0.39 & 51.5 & 0.80$\times$ \\
    512 & 0.88 & 0.94 & 41.8 & 1.07$\times$ & 0.81 & 44.8 & 0.92$\times$ & 0.99 & 51.5 & 1.12$\times$ & 0.89 & 55.5 & 1.01$\times$ \\
    1024 & 1.05 & 4.06 & 71.0 & 3.86$\times$ & 3.22 & 83.0 & 3.07$\times$ & 5.17 & 110.0 & 4.92$\times$ & 4.63 & 126.0 & 4.41$\times$ \\
    2048 & 1.10 & 7.34 & 188.0 & 6.69$\times$ & 5.76 & 236.0 & 5.25$\times$ & 16.66 & 344.0 & 15.19$\times$ & 14.69 & 408.0 & 13.39$\times$ \\
    4096 & 1.10 & 9.83 & 656.0 & 8.94$\times$ & 7.59 & 848.0 & 6.90$\times$ & 35.35 & 1280.0 & 32.16$\times$ & 30.82 & 1536.0 & 28.03$\times$ \\
    8192 & 1.10 & 11.01 & 2560.0 & 10.00$\times$ & 8.45 & 3328.0 & 7.67$\times$ & 58.73 & 5120.0 & 53.32$\times$ & 50.62 & 6144.0 & 45.95$\times$ \\ 
        16384 & 1.10 & 11.58 & 10240.1 & 10.50$\times$ & 8.87 & 13312.1 & 8.05$\times$ & 75.67 & 20480.1 & 68.65$\times$ & 63.75 & 24576.1 & 57.84$\times$ \\
    24576 & 1.10 & 11.77 & 23040.1 & 10.67$\times$ & 9.01 & 29952.1 & 8.17$\times$ & 84.30 & 46080.1 & 76.44$\times$ & 71.17 & 55296.1 & 64.53$\times$ \\
    32768 & 1.10 & 11.88 & 40960.1 & 10.77$\times$ & 9.09 & 53248.1 & 8.24$\times$ & 88.08 & 81920.1 & 79.85$\times$ & 74.19 & 98304.1 & 67.26$\times$ \\
    \bottomrule
  \end{tabular}
\end{table*}

\section{Robustness of standalone emulated \ac{GEMM}}
\label{sec:robustness}

\subsection{Is Ozaki reproducible?}
 A common misconception about emulation is that emulated results are {\it not bit-wise reproducible}: the results of two emulated \ac{GEMM} operations might not be the same for the same inputs; this is a valid concern as even small variations in the results can lead to unstable behavior in numerically sensitive applications. The published literature clearly disproves this~\cite{uchino2026error, mukunoki2020dgemm} and both Ozaki schemes are reproducible by design. 

\begin{figure*}
    \centering
     \centering
    \begin{subfigure}[t]{0.5\textwidth}
        \centering
        \includegraphics[width= \textwidth]{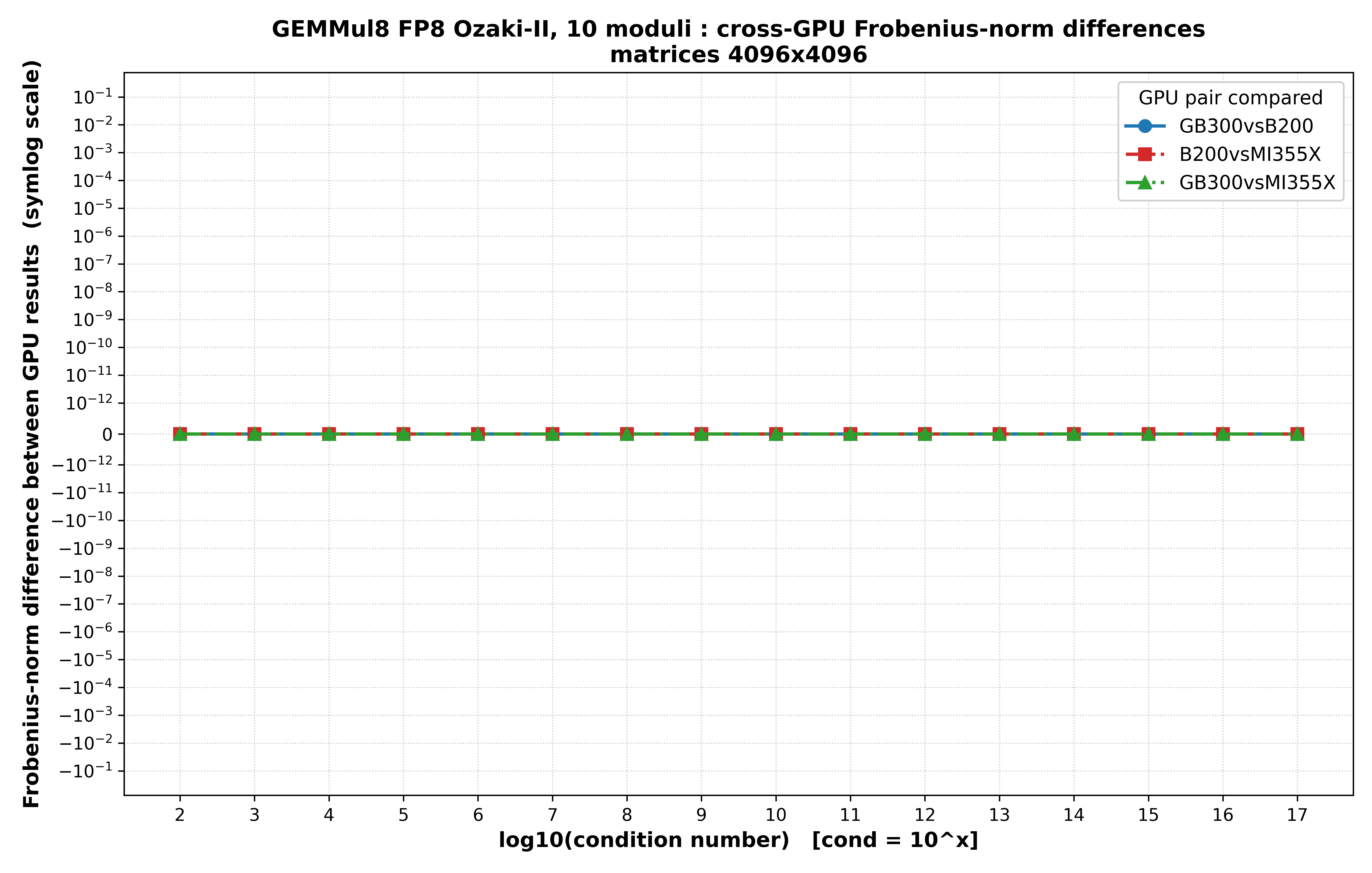}
        \caption{FP8 Ozaki II-10}
    \end{subfigure}%
    ~ 
    \begin{subfigure}[t]{0.5\textwidth}
        \centering
        \includegraphics[width= \textwidth]{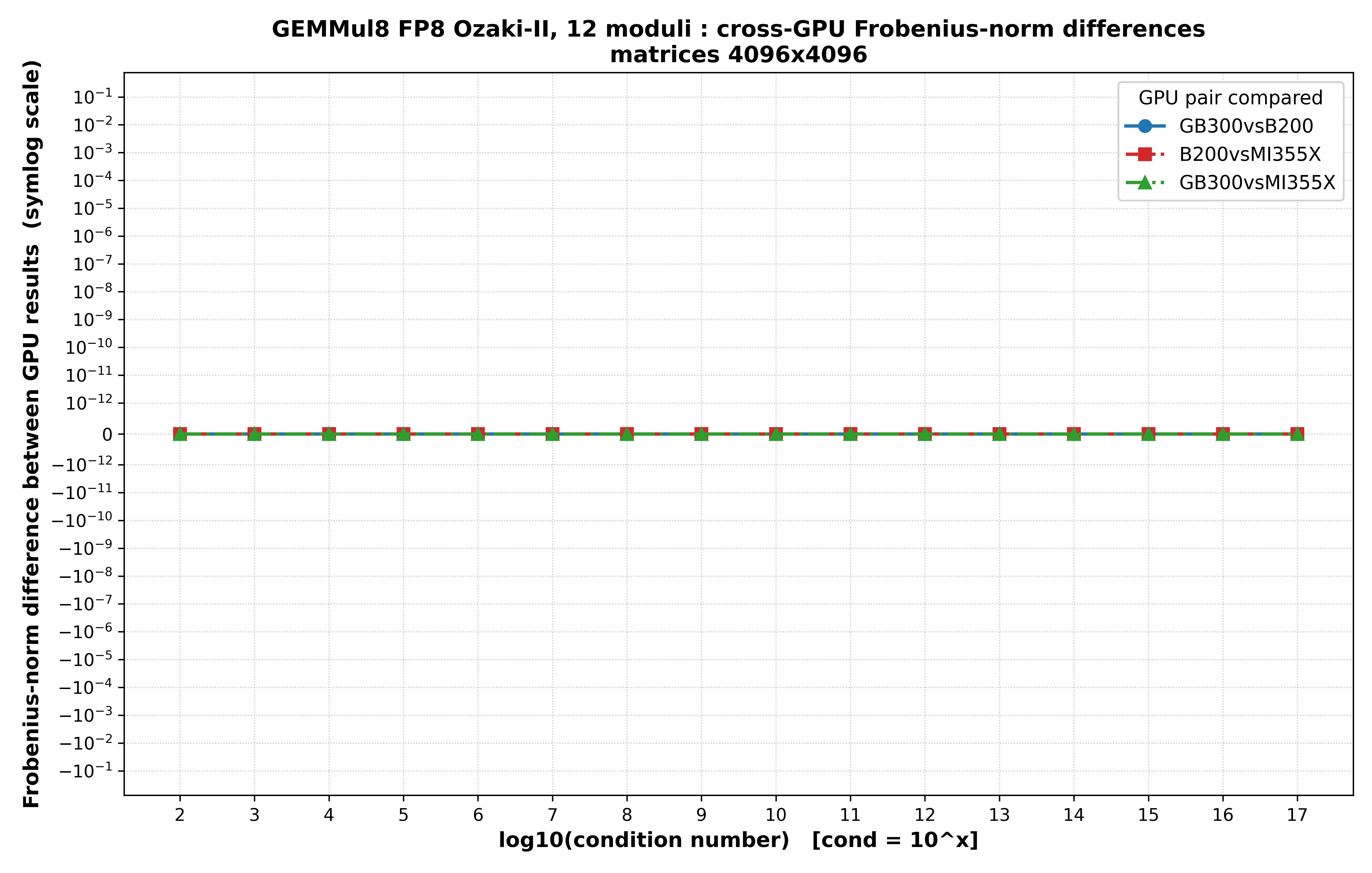}
        \caption{FP8 Ozaki II-12}
    \end{subfigure}
        \begin{subfigure}[t]{0.5\textwidth}
        \centering
        \includegraphics[width= \textwidth]{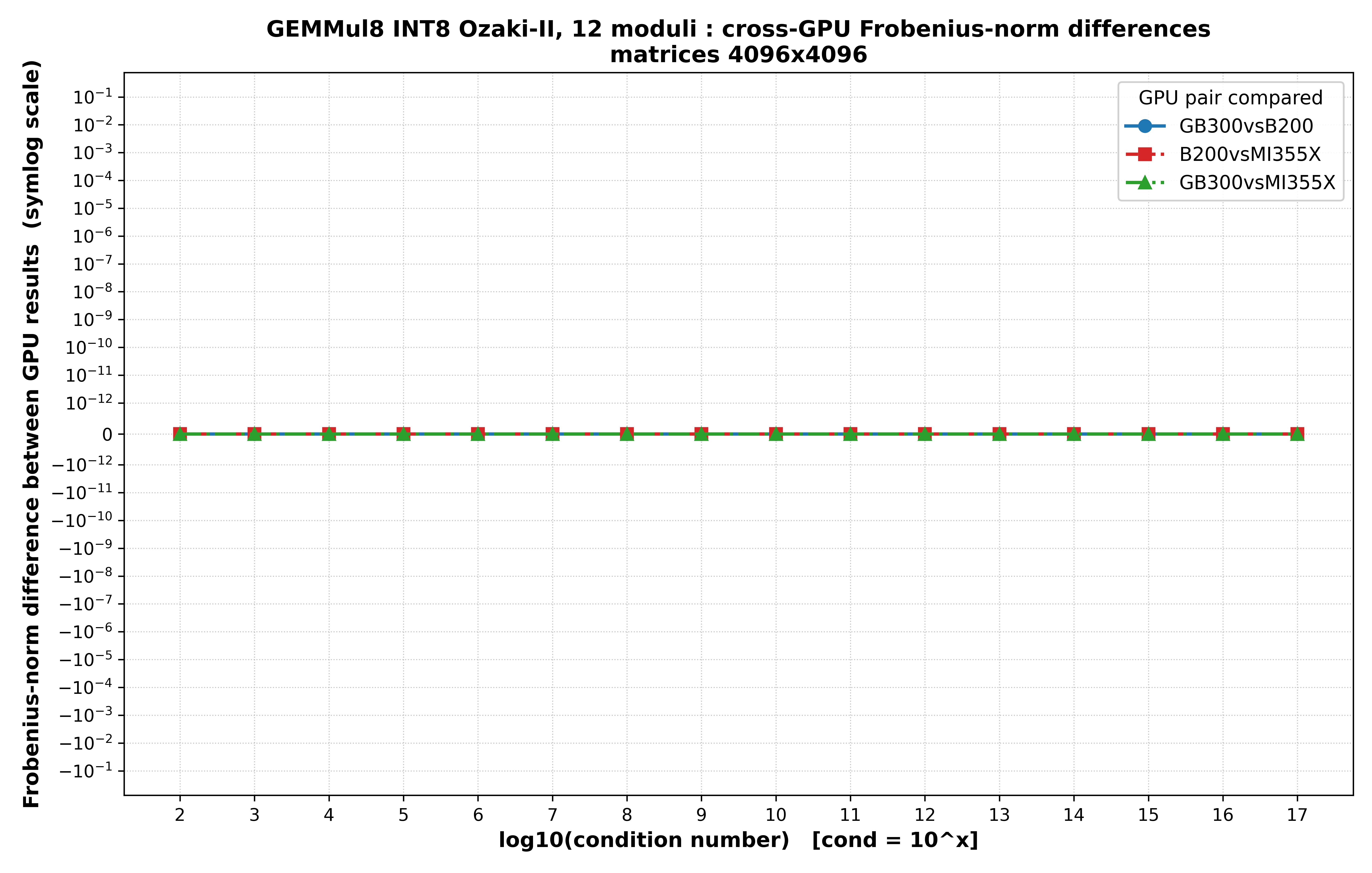}
        \caption{INT8 Ozaki II-12}
    \end{subfigure}\begin{subfigure}[t]{0.5\textwidth}
        \centering
        \includegraphics[width= \textwidth]{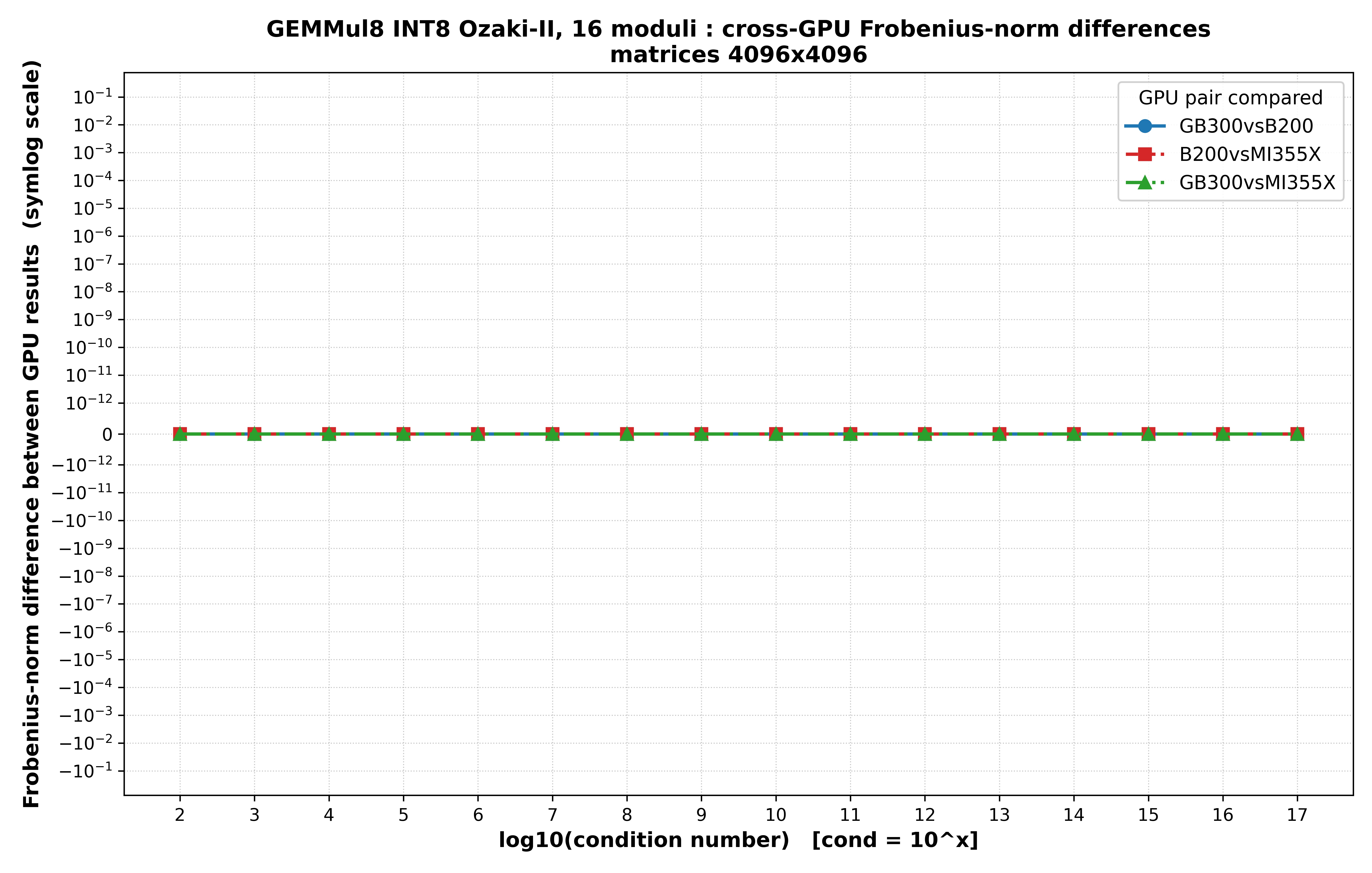}
        \caption{INT8 Ozaki II-16}
    \end{subfigure}
    \begin{subfigure}[t]{0.5\textwidth}
        \centering
        \includegraphics[width= \textwidth]{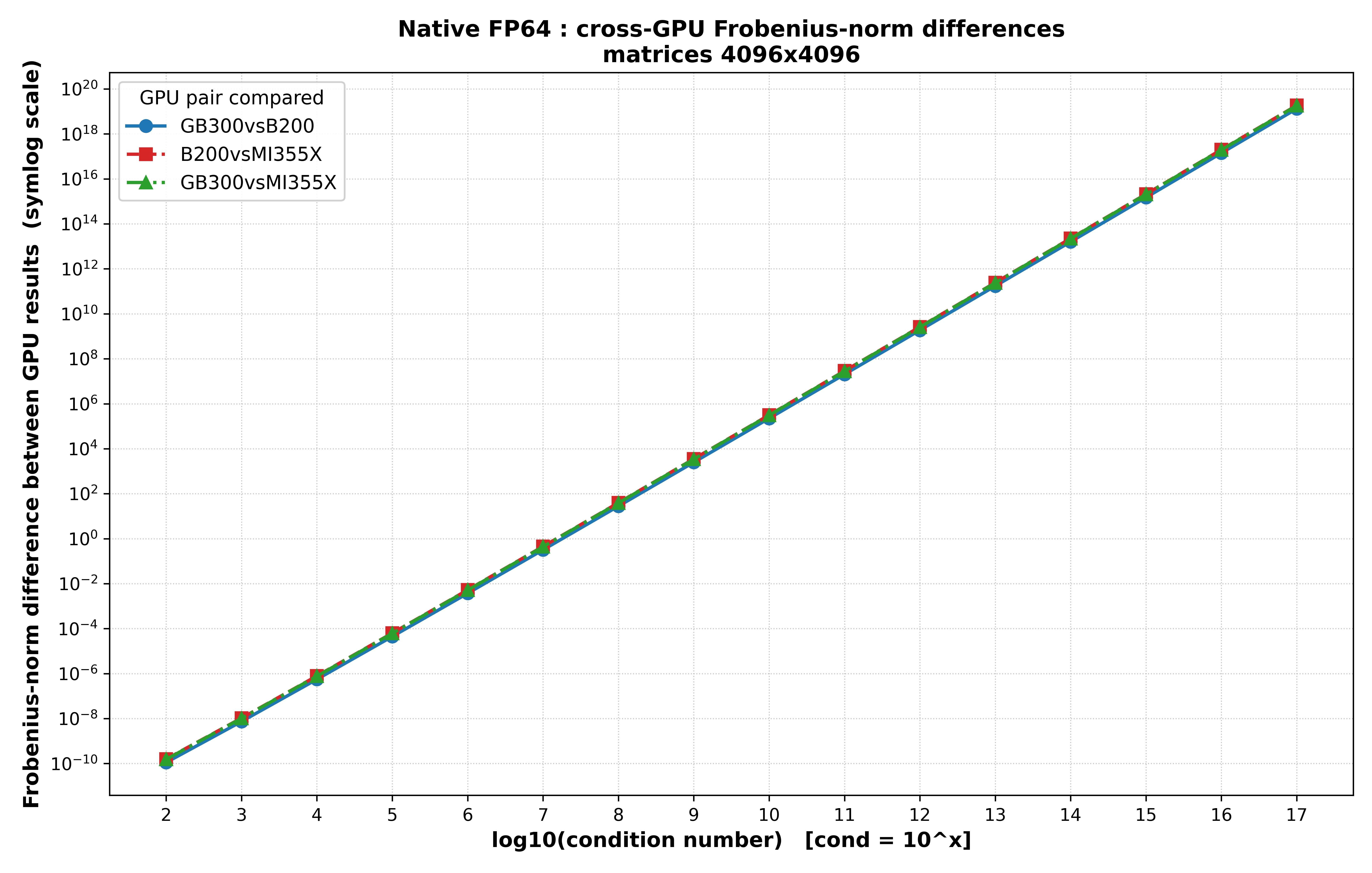}
        \caption{Native FP64}
    \end{subfigure}
    \caption{Comparison of results across three GPUs
(\mbox{AMD MI355X}, \mbox{NVIDIA B200}, and \mbox{NVIDIA GB300}).}
    \label{fig:3waycomp}
\end{figure*}

 To confirm reproducibility in practice, we run the following experiment: we generate a set of $4096 \times 4096$ matrices with prescribed condition numbers. We save the matrices and multiply them using  INT8 Ozaki-II 12, INT8 Ozaki-II 16, FP8 Ozaki-II 10,  FP8 Ozaki-II 12, and FP64 on three compute systems from Table~\ref{tab:systems}. We save the computed $C$s (five results per matrix pair) and compare them per method. The final results are shown in Figure~\ref{fig:3waycomp}. In all cases of Ozaki II scheme, the results are identical (note here Frobenius norm is not scaled). The results of native FP64 differ for the native \ac{GEMM} and the difference grows with the growth of the condition number. We conclude that not only Ozaki  scheme II produces bit-wise reproducible results: it is actually more numerically reliable than the native \ac{GEMM}, see Figure~\ref{fig:3waycomp}. The results are identical, regardless of the platform and number of moduli used. It is worth emphasizing that it does not mean that Ozaki scheme is "more correct" (closer to the analytical true solution): it simply means that for the same input we can expect exactly the same output (correct or not, this test does not check correctness); this is not the case for FP64 computations for high condition numbers, see Figure~\ref{fig:3waycomp}, subplot (e).

 \subsection{Corner cases}
\ac{GPU}-accelerated \ac{BLAS} libraries such as ROCm roc\ac{BLAS}~\cite{rocblas2026} perform large GEMMs by splitting the work into tiles on several levels; bottom level tile sizes are aligned with the sizes needed for hardware supported \ac{MFMA} operations. These operations act on chunks of data with shapes such as, for example, 16x16x4 ($16 \times 4$ matrix multiplied by $4 \times 16$ matrix) for MI355X FP64 \ac{MFMA} and 32x32x32 ($32 \times 32$ matrix multiplied by $32 \times 32$ matrix) for MI355X INT8 \ac{MFMA}. When the matrix size is not aligned with the \ac{MFMA} size, more tiles are launched to cover full matrix multiplication. Hence, we expect a slight performance drop when using misaligned matrix sizes such as $4097 \times 4097$. 

\begin{figure*}
    \centering
    \includegraphics[width=\linewidth]{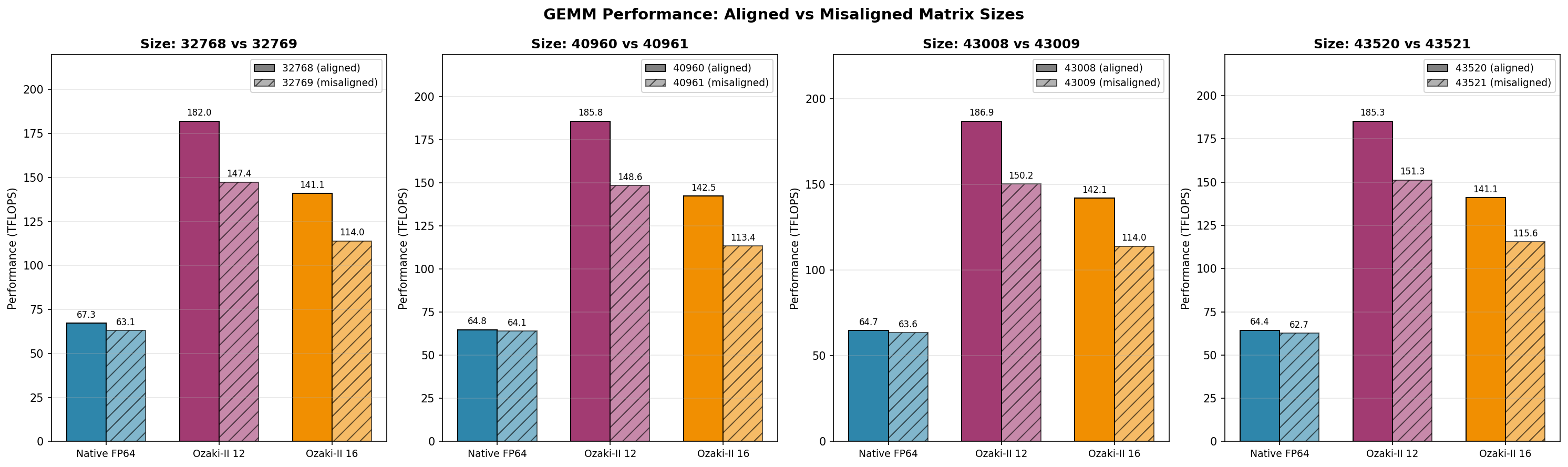}
    \caption{GEMMul8 on AMD MI355X: influence of misaligned matrix sizes on performance}
    \label{fig:misaligned_sizes}
\end{figure*}

We note that Ozaki II performs 12 or 16 GEMMs instead of one, so we expect performance to drop further as the effects of misaligned matrix size accumulate, see Figure~\ref{fig:misaligned_sizes}. We can expect this to be worse for Ozaki I as it executes more GEMMs than Ozaki II. In the figure the performance drops less than 5\% for FP64 \ac{GEMM} and can drop more than 25\% for the emulated GEMMs.

\section{Conclusions and future work}
\label{sec:conclusions}
Our results demonstrate that native FP64 remains an essential
architectural capability for \ac{HPC}. The roofline analysis shows that FP64 performance, not only memory bandwidth, determines
application performance. Reducing a device's ratio of FP64
throughput to memory bandwidth lowers its roofline ridge point, causing kernels that are bandwidth-bound on current \ac{HPC} \ac{GPU}s to become compute-bound on future \ac{AI}-specialized accelerators.

\ac{GEMM} emulation cannot address this gap broadly. In our dataset, ten of the eleven applications and benchmarks have no dominant \ac{DGEMM} kernel and therefore depend primarily on native vector FP64 performance. Even when a workload is amenable to Ozaki-style emulation, its practical benefit is constrained by accuracy,workspace, problem size, and integration overheads. Accelerators such as Vera Rubin and MI455X may be highly effective for \ac{AI}, but their reduced native FP64 throughput prevents them from serving as general replacements for \ac{HPC}-oriented \ac{GPU}s across this workload portfolio. These results motivate continued investment in science-focused accelerators such as MI430X, which preserves substantial hardware-based FP64 throughput while also supporting modern \ac{AI} workloads.

Future work will extend this study from standalone \ac{GEMM} and selected application integrations to a broader set of production \ac{HPC} workloads. We will investigate how the observed accuracy, scaling, and workspace sensitivities propagate through complete solvers and multi-node runs. We will also explore whether \ac{AI}-for-science surrogates and other data-driven models can safely exploit reduced- and mixed-precision
arithmetic while preserving the accuracy required by scientific
workflows. When MI430X becomes available in 2027, we plan to validate our roofline projections and compare native high-throughput FP64 execution, mixed-precision approaches, and Ozaki-style emulation on the same architectures.

\ac{AI}-for-science surrogate models provide another complementary direction, potentially replacing selected high-cost solver components or enabling reduced- and mixed-precision execution while retaining a conventional high-fidelity correction mechanism~\cite{obiolssales2020cfdnet}.

\section{Reproducibility}

All results in this work are meant to be reproduced. We already provided details of hardware and compilers, and the commit hashes of the open source repositories used in this work. {\color{black} The code used in this paper can be found in this repository: \href{https://github.com/kswirydo/ozaki_tests/}{https://github.com/kswirydo/ozaki\_tests/}}

\bibliographystyle{IEEEtran}
\bibliography{
bib/ref_ks
}

\end{document}